\documentclass{article}

\usepackage[table,xcdraw]{xcolor}
\usepackage{arxiv}

\usepackage[utf8]{inputenc} 
\usepackage[T1]{fontenc}    
\usepackage{hyperref}       
\usepackage{url}            
\usepackage{booktabs}       
\usepackage{amsfonts}       
\usepackage{nicefrac}       
\usepackage{microtype}      
\usepackage{lipsum}
\usepackage{graphicx}
\graphicspath{ {./images/} }
\usepackage{multirow}
\usepackage{amsthm}          
\usepackage[tbtags]{amsmath}
\usepackage{amssymb}
\usepackage{booktabs}
\usepackage{makecell}
\usepackage{subcaption}

\usepackage{natbib}
\title{Hurricane Impact Index for Assessing Direct and Indirect Hazards in Central America}

\author{
 Manrique Camacho\\
  Centro de Investigaciones Geofísicas\\
  Escuela de Estadística\\
  Universidad de Costa Rica\\
  San José, Costa Rica \\
  \texttt{manrique.camacho@ucr.ac.cr} \\  
  \And
  Amanda Cedeño\\
  Centro de Investigaciones Geofísicas\\
  Escuela de Estadística\\
  Universidad de Costa Rica\\
  San José, Costa Rica \\
  \texttt{amanda.cedeno@ucr.ac.cr} \\  
   \And
 Luis A. Barboza \\
  Centro de Investigación en Matemática Pura y Aplicada\\
  Centro de Investigaciones Geofísicas\\
  Escuela de Matemática \\
  Universidad de Costa Rica\\
  San José, Costa Rica \\
  \texttt{luisalberto.barboza@ucr.ac.cr} \\
  \And
  Shu Wei Chou-Chen \\
  Centro de Investigación en Matemática Pura y Aplicada\\
  Centro de Investigaciones Geofísicas\\
  Escuela de Estadística \\
  Universidad de Costa Rica\\
  San José, Costa Rica \\
  \texttt{shuwei.chou@ucr.ac.cr} \\
  \And
  Mario J. Gómez \\
  Centro de Investigaciones Geofísicas\\
  Universidad de Costa Rica\\
  San José, Costa Rica \\
  \texttt{mariojavier.gomez@ucr.ac.cr} \\
  \And
  Hugo G. Hidalgo \\
  Centro de Investigaciones Geofísicas\\
  Escuela de Física \\
  Universidad de Costa Rica\\
  San José, Costa Rica \\
  \texttt{hugo.hidalgo@ucr.ac.cr} \\
}

\begin{document}
	
\maketitle
\begin{abstract}
Hurricanes rank among the most destructive natural hazards. They are complex phenomena that can cause both direct damage along their path and indirect impacts due to heavy rainfall and strong winds, with effects varying according to regional topography. In this paper, we propose a Hurricane Impact Index to assess both direct and indirect hazards, and we demonstrate its applicability to the Central American region. The index is constructed so that we can decompose these effects across multiple dimensions of time and space, enabling a detailed analysis of the intensity and distribution of hurricane impacts. 
\end{abstract}

\keywords{Tropical Cyclone, Hurricane Impact Index, Vorticity, Wind Speed, Precipitation, Orography,  Natural Hazards, Central America}

\section{Introduction}

Hurricanes are among the most destructive natural hazards, causing significant human and material losses, particularly in regions such as Central America, North America, and the Caribbean \citep{martinez2023assessment, nordhaus2006economics, sivakumar2005impacts, quesada2024assessing}. These events are complex phenomena involving multiple interacting climatic processes beyond mere wind speed \citep{Mehdi-2014,Alipour-2022,Sammonds-2023}. Given their far-reaching consequences—particularly in the context of rising global temperatures, which are expected to continue increasing and influencing hurricane risk \citep{emanuel2021response, little2015joint, dinan2017projected, gori2022tropical, knutson2020tropical, woodruff2013coastal}—it is essential to develop reliable methods for comprehensively quantifying their effects.

The low-pressure centers where hurricanes form in adjacent seas and oceans generate both direct and indirect consequences for surrounding areas \citep{perez2016distribucion, hidalgo2020identification, pena2002characteristics, alfaro2014analisis}. A cyclone does not always need to make direct landfall to influence a region; its strategic positioning can indirectly affect distant communities \citep{hidalgo2020identification, hidalgo2023probability}. Southern Central America exemplifies this situation: although Panama and Costa Rica are rarely struck directly, past events have demonstrated that indirect effects can still cause severe disruptions when low-pressure centers are located just a few hundred kilometers away \citep{hidalgo2020identification,hidalgo2023probability,pena2002characteristics}.

While previous studies have proposed multi-hazard indices to measure the effects of tropical cyclones \citep{Qi-2018,Bloemendaal-2021,Alipour-2022,Song-2024,Tripathy-2024}, most have not explicitly distinguished between direct and indirect influences. This distinction is crucial for improving assessments and disaster preparedness. To address this gap, this study introduces a novel Hurricane Impact Index (HII) designed to quantify both direct and indirect effects, thereby offering a more comprehensive approach to these phenomena. In this study, we use the term Hurricane to describe our metric; however, it is important to note that the same index could be referred to as the Cyclone Impact Index (CII) or Typhoon Impact Index (TII), depending on the region of interest. These different names reflect regional naming conventions for tropical cyclones but refer to the same underlying concept and methodology.

The proposed index builds upon the Coastal Hurricane Frequency (CHF) index developed by Balaguru et al. (\citeyear{balaguru2023increased}), but it also incorporates new key variables, such as wind speed and precipitation, since the potential for damage caused by hurricanes is largely determined by these factors \citep{paul2009relatively, seo2017tropical, peduzzi2012global}. Moreover, the index accounts for vorticity and orographic influences, as mountainous regions tend to experience heavy rainfall during hurricanes due to terrain effects \citep{smith2009orographic, zachry2015national}. This scheme enhances the precision of hurricane effect measurement. Additionally, validation using events from 2016 to 2022 confirms its applicability and reliability across diverse geographic and climatic conditions.

This research advances the understanding of hurricane-related hazards by offering a robust and adaptable tool for assessment. The findings of this study hold significance for policymakers, disaster management authorities, and the scientific community, as they contribute to improved planning and response strategies for future events.

\section{Methods}

\subsection{Data}\label{Data}

To develop the HII, a combination of meteorological, geospatial, and disaster impact data were used. The selected datasets provided key variables necessary to quantify hurricane intensity, structural dynamics, and observed effects. These variables included hurricane trajectories, translational speed, wind speed, precipitation, vorticity, and wind vectors, as well as orographic data to account for terrain influence on rainfall distribution. Additionally, historical disaster records were incorporated to validate the proposed index against real-world impacts. The study area for the analysis covered the Central American isthmus, including parts of Mexico (specifically the regions bordering the isthmus), Guatemala, Belize, Honduras, El Salvador, Nicaragua, Costa Rica and Panama, as illustrated in Fig. \ref{fig:adminplot}. This region was selected due to its geographic exposure to both the direct and indirect effects of tropical cyclones, as highlighted in previous studies \citep{hidalgo2020identification,hidalgo2023probability,pena2002characteristics}. Some areas within the region frequently experience direct landfalls, while others, particularly in the southern part of the isthmus, are more commonly affected by indirect impacts from nearby low-pressure systems. The region’s complex topography and diverse climatic conditions also provide an ideal context for evaluating the sensitivity and robustness of the proposed methodology across a range of environmental scenarios. The following data sources were utilized:

\subsubsection{HURDAT2}

The second-generation HURricane DATa (HURDAT2) dataset provides six-hourly records of tropical and subtropical cyclones in the Atlantic Ocean \citep{landsea2013atlantic, nhc_data_archive}. Managed by the National Hurricane Center (NHC) under the United States National Oceanic and Atmospheric Administration (NOAA), it contains information on location, wind speed, and central pressure for all known events from 1851 to the present. We use this dataset to extract the trajectories of the hurricanes of interest and to calculate their translational speeds at specific points.

\subsubsection{ERA 5}

ERA5 is the fifth-generation reanalysis dataset from the European Centre for Medium-Range Weather Forecasts (ECMWF). It provides hourly, daily, and monthly climate data spanning from 1940 to the present \citep{hersbach2020era5,ecmwf_reanalysis_v5}. For this study, we use ERA5 variables at a spatial resolution of $0.25^\circ$. Precipitation and wind speed were extracted for the past 30 years to support historical records. Wind vectors were computed by combining the 10m u-component (zonal) and 10m v-component (meridional) of wind. The direction of each vector was calculated using the arctangent of the ratio between components, following the formula $\Theta = \tan^{-1} \left( \frac{A_y}{A_x} \right)$, which yields the angle in radians. For vorticity, we consider pressure levels at 850 hPa and 200 hPa. The variable is computed as the average between these levels, representing the lower and upper bounds of the atmospheric column where cyclonic dynamics are most pronounced. This multi-level averaging approach provides a more robust representation of the hurricane’s structural coherence and dynamical influence \citep{fischer2022analysis}.

\subsubsection{AppEEARS}

The HII implicitly incorporates angular constraints imposed by the orientation of wind vectors relative to surrounding mountain ranges. To define these angular constraints, we analyze orographic data from the Application for Extracting and Exploring Analysis Ready Samples (AppEEARS), a web application developed by the National Aeronautics and Space Administration (NASA) for accessing and transforming geospatial data \citep{appeears}. Specifically, we use elevation data from the ASTER Global Digital Elevation Model (GDEM), which provides the topographic detail necessary to assess terrain influence on wind flow.

\subsubsection{DesInventar}

DesInventar is an inventory system designed for collecting, querying, and analyzing disaster-related data. Sponsored by the United Nations Office for Disaster Risk Reduction (UNDRR) as part of the Sendai Framework, it contains both quantitative and qualitative information on human, infrastructural, and economic losses for participating countries \citep{desinventar}. We employ the DataCards variable as a proxy for overall hurricane damage; it records the number of disaster events (e.g., floods, storms, storm surges) attributed to each storm.

\subsection{Statistical Methods}

As an initial step in constructing our indicator, we partitioned the study area into 1°~$\times$~1° latitude--longitude grid cells to balance its spatial bias and variance. We then aggregated all atmospheric variables—originally available at finer resolutions—by computing their mean within each grid cell, ensuring a consistent spatial scale across datasets. The temporal resolution was set to six-hour intervals, matching the HURDAT2 dataset. These design choices facilitate a systematic analysis of spatiotemporal patterns and the evaluation of hurricane impacts.

We then computed the impact index for each hurricane event, a procedure that effectively captured spatial and temporal variations in hurricane effects. Our analysis focused on six major hurricanes between 2016 and 2022: Otto (2016), Nate (2017), Eta (2020), Iota (2020), Bonnie (2022), and Julia (2022). This time frame was chosen to illustrate recent trends in extreme weather driven by climate variability and the pronounced coastal activity along Nicaragua and Costa Rica.

We defined the impact index by integrating a direct‐impact component that quantifies damage along the hurricane trajectory. Following \citet{balaguru2023increased}, we recorded hurricane positions at six‐hour intervals and incorporated their translational speed. We delineated directly affected grid cells using an indicator function that activates when a cell lies within the storm’s spatial influence, determined by vorticity—a key diagnostic of cyclonic structure, intensity, and organization \citep{fischer2022analysis}. We defined the direct‐impact region as those cells exceeding the hurricane‐specific 90th percentile of vorticity and located within 500 km of the cyclone center, corresponding to a $5\times5$ grid at 1° resolution \citep{khouakhi2017contribution}. This approach maintains spatial precision and remains physically grounded in the dynamical characteristics of each event.

To integrate precipitation and wind speed into the direct-impact component, we defined a variable-specific weighting factor. Let $x$ denote an atmospheric variable in a given grid cell; we normalize it as

\begin{equation}
	x_{z} = \frac{[x - \text{P}_{10}(x)]_{+}}{\text{P}_{90}(x) - \text{P}_{10}(x)},
	\label{eq:scaled_variable}
\end{equation}

\noindent
where $P_p(x)$ is the $p$th percentile of $x$ computed from the cell’s historical record and $[\cdot]_+$ denotes the positive-part operator. We employ the 10th percentile rather than the minimum to prevent anomalously low baselines from inflating normalized values, thereby highlighting only measurements that exceed a typical low-impact threshold. The denominator, $P_{90}(x)-P_{10}(x)$, provides a robust dispersion metric that resists distortion by outliers. Negative numerators are set to zero so that sub-threshold values represent neutral, rather than negative, effects. To accommodate seasonality, percentiles are calculated using historical data for the event’s month; if a hurricane spans two months, normalization is applied separately for each. Finally, we apply this weighting to the product of normalized precipitation and wind speed over the 1993–2023 reference period (see equation \eqref{eq:HII_direct}), thereby capturing how heavy rainfall amplifies hydrological damage and strong winds exacerbate structural destruction, up to the complete isolation of affected areas.

Consequently, the direct effect of a hurricane $h$, at grid cell $i$ and time $t$ is defined as follows:

\begin{equation}
	{DE}_{i,h,t} = 
	{D_{i,h,t}\cdot\frac{\ell}{\Delta\cdot{v_{h,t}}}\cdot{{(W_{i,t}\cdot P_{i,t})_z}}},
	\label{eq:HII_direct}
\end{equation}

\noindent
where

\begin{itemize}
	\item[-] $D_{i,h,t}$: Indicator function with $1$ if hurricane $h$ is located within grid $i$ at time $t$ and meets the vorticity region conditions described above, and $0$ otherwise.
	\item[-] $\ell$: Length of grid $i$.
	\item[-] $\Delta$: Timestep $= 6$ hours (according to HURDAT2's time resolution).
	\item[-] $v_{h,t}$: Translational speed of hurricane $h$ at time $t$.
	\item[-] $W_{i,t}$: Wind speed within grid $i$ at time $t$.
	\item[-] $P_{i,t}$: Precipitation within grid $i$ at time $t$.
	\item[-] $(\cdot)_z$: Indicates that the variable $(\cdot)$ is scaled using the equation \eqref{eq:scaled_variable} and the historical records.
\end{itemize}

\begin{figure}
	\centering
	\begin{subfigure}[b]{0.48\textwidth}
		\includegraphics[scale = 0.4]{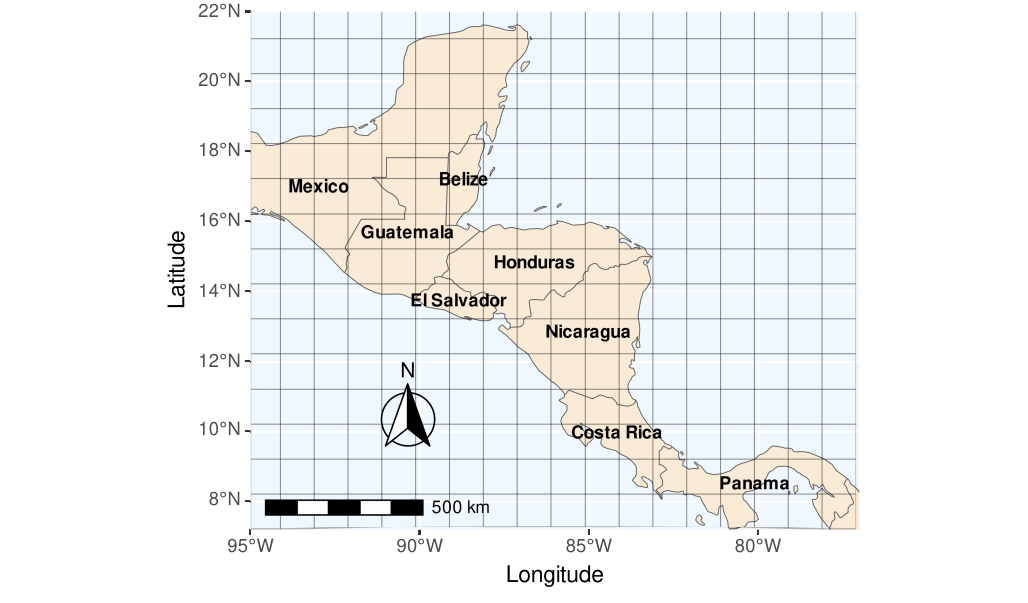}
		\caption{}
		\label{fig:adminplot}
	\end{subfigure}
	\begin{subfigure}[b]{0.48\textwidth}
		\includegraphics[scale = 0.4]{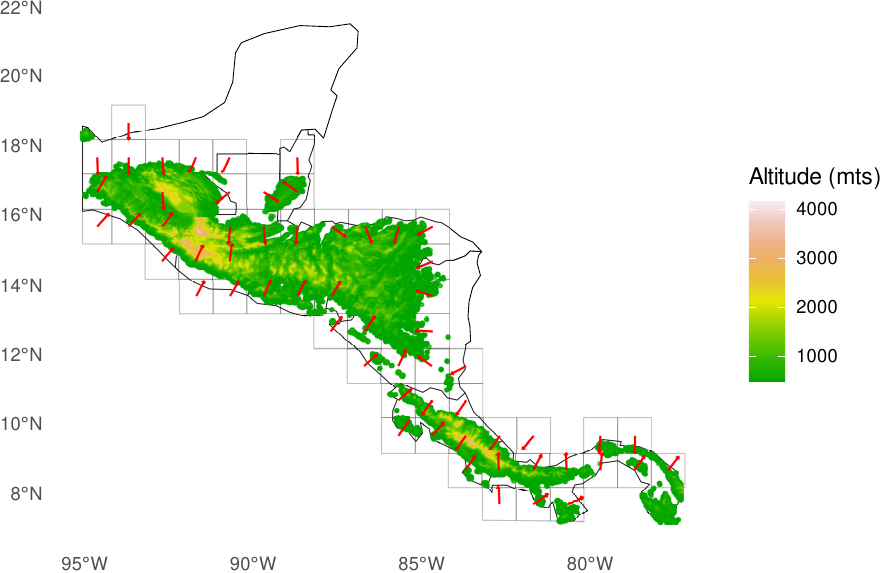}
		\caption{}
		\label{fig:topoplot}
	\end{subfigure}
	\caption{Panel a: Area of study and grid definition, using the resolution of 1°~$\times$~1° latitude--longitude cells. Panel b: Topographical data (above 500 meters) using AppEEARS, combined with the proposed angular mean restrictions.}
\end{figure}

While the direct impact quantifies damage along the hurricane’s trajectory, the indirect impact captures broader atmospheric influences that extend beyond this path, shaped by dynamic airflow and regional topography. We operationalize indirect effects with a binary indicator function that assigns a value of 1 to grid cells experiencing secondary mechanisms—such as wind advection or pressure‐gradient forcing. To link these mechanisms to orographic modulation, we defined angular influence limits for each cell using AppEEARS topographical data (mountains above 500 m). Specifically, we constrained wind‐vector directions to within ±45° of the principal axis of the nearest mountain range—itself oriented near 90° relative to the Pacific coastline—to identify cells under maximal indirect influence due to possible orographic lifting (see Fig. \ref{fig:topoplot}). A cell is marked indirectly affected if its wind vector falls within these angular bounds. This parameterization reflects the role of mountain chains in channeling or disrupting atmospheric flow, thereby governing the spatial extent and orientation of cyclone‐induced secondary effects. Consistent with prior studies \citep{pena2002characteristics,hidalgo2020identification}, which demonstrate that Caribbean cyclones generate low‐pressure centers that attract humid Pacific air over mountain slopes by orographic lifting, our analysis reveals that, despite Caribbean origins, the most pronounced indirect impacts in southern Central America occur along the Pacific coast.

To extend the CHF index proposed by \cite{balaguru2023increased} to address the indirect effects in the Central American region, we incorporated precipitation as a weighting factor for the indirect impact component, given that higher precipitation levels tend to amplify these effects. Consequently, we define the indirect effect of a hurricane $h$ at grid cell $i$ and time $t$ as follows:
\begin{equation}
	{IE}_{i,h,t} = (1-D_{i,h,t})\cdot I_{i,h,t}\cdot\frac{\ell}{\Delta\cdot{v_{h,t}}} \cdot{{{(P_{i,t}})_z}},
	\label{eq:HII_indirect}
\end{equation}
where
\begin{itemize}
	\item[-] $D_{i,h,t}$: Indicator function with $1$ if hurricane $h$ is located within grid $i$ at time $t$ and meets the vorticity region conditions, 0 otherwise.
	\item[-] $\ell$: Length of grid $i$.
	\item[-] $\Delta$: Timestep $= 6$ hours (according to HURDAT2's time resolution).
	\item[-] $v_{h,t}$: Translational speed of hurricane $h$ at time $t$.
	\item[-] $I_{i,h,t}$: Indicator function, equals 1 if the wind vector at grid cell $i$ and time $t$ falls within its angular influence zone for hurricane $h$, 0 otherwise.
	\item[-] $P_{i,t}$: Precipitation within grid $i$ at time $t$.
	\item[-] $(\cdot)_z$: Indicates that the variable $(\cdot)$ is scaled using the equation \eqref{eq:scaled_variable}, from the historical records.
\end{itemize}

By aggregating and standardizing both direct and indirect effects over the hurricane’s lifetime, we define the grid‐level Hurricane Impact Index (HII) as
\begin{equation}
	{\text{HII}}_{i,h} = 
	\left[ \left(\sum_{t \in T_h}{DE}_{i,h,t}\right)_{s} + 
	\left(\sum_{t \in T_h}{IE}_{i,h,t}\right)_{s} \right]_{s}
	\label{eq:HII_1}
\end{equation}
where $T_h$ denotes the set of six‐hour time points during which hurricane $h$ was active in the HURDAT2 database. Each summed component is first normalized according to
\begin{equation}
	(x)_{s} = \frac{x - \text{min}(x)}{\text{max}(x) - \text{min}(x)} 
	\label{eq:scaled_variable2}
\end{equation}
ensuring that the individual direct and indirect effect scores lie within $[0,1]$. A second application of the same normalization to their sum then constrains the combined HII to $[0,1]$. This sequential normalization preserves scale consistency, thereby facilitating the interpretation and visualization of the index.

Also, the decomposition of direct and indirect effects leads to different mathematical expressions that provide a deeper understanding of the underlying dynamics. By separating these effects, it is possible to assess their individual contributions and interactions, allowing for a more detailed interpretation of the overall impact. The following expressions illustrate how these components are formulated and analyzed across different approaches, providing structured framework for quantifying their influence.

By aggregating our index across a defined set of hurricanes, we capture the cumulative impact of multiple storms on the region of interest, quantify overall exposure, and identify damage hotspots. This approach also reveals spatial vulnerability patterns, highlighting areas with elevated susceptibility to hurricane-induced effects. The resulting stacked index for grid cell $i$, denoted $\mathrm{HII}_{R,i}$, is therefore defined as follows:

\begin{equation}
	\text{HII}_{R,i} = \left( \sum_{h \in H} \text{HII}_{i,h} \right)_s
	\label{eq:HII_2}
\end{equation}
where $H$ is the set of hurricanes under study.     

Both effects in Equations \eqref{eq:HII_direct} and \eqref{eq:HII_indirect} allow us to aggregate grid cell data and visualize the temporal evolution of a specific hurricane’s impact. As an initial indicator, we aggregate the direct and indirect effects across all grid cells in the spatial domain \(I\), where the hurricanes under study occur. This approach preserves the inherent magnitude of each effect, providing a clear representation of their evolution on the original scale throughout the event. By displaying the time series in its unaltered form, the formula offers straightforward insight into the day-by-day variations of each effect, enabling direct comparison of their absolute contributions during the hurricane:

\begin{equation}
	\sum_{i \in I} {DE}_{i,h,t} + \sum_{i \in I} {IE}_{i,h,t}
	\label{eq:HII_5}
\end{equation}

Additionally, if we compare the sum in Equation \eqref{eq:HII_5} with the total number of grid cells \(n\) that could potentially exhibit a positive HII within the region of interest, we obtain an approximation of the proportion of cells displaying some type of effect—whether direct or indirect—at any given time. For ease of comparison, \(n\) is considered constant with respect to time and across all hurricanes under study. We denote this modified index as HII\(_{TS1}\):

\begin{equation}
	\text{HII}_{\text{TS1},h,t} = \left[\left(\sum_{i \in I} DE_{i,h,t}\right) +  \left(\sum_{i \in I} IE_{i,h,t}\right)\right] \cdot \frac{100}{n}
	\label{eq:HII_3}
\end{equation}

A second modification of the temporal index, denoted as \(\text{HII}_{\text{TS3}}\), adopts a similar framework but implements an alternative normalization strategy. Rather than applying a uniform scaling factor of \(\frac{100}{n}\) to the combined components, we normalize each component individually via a min-max transformation to a range of \(0\) to \(1\). Consequently, when the two normalized components are summed, the total can vary between \(0\) and \(2\). This method facilitates a more precise identification of the day during the hurricane when the effect—whether direct or indirect—was most pronounced:

\begin{equation}
	\text{HII}_{\text{TS2},h,t} = \left(\sum_{i \in I} {DE}_{i,h, t}\right)_s+\left(\sum_{i \in I} {IE}_{i,h,t}\right)_s
	\label{eq:HII_4}
\end{equation}

\section{Results}

The analysis of the HII reveals a comparative overview of the direct and indirect effects for each hurricane under study: Otto (2016), Nate (2017), Julia (2022), Bonnie (2022), Iota (2020), and Eta (2020). The HII scale, ranging from 0 to 1, provides a quantifiable measure of impact based on the cumulative impacts recorded within each grid cell, segmented into direct and indirect categories. 

\begin{figure}
	\centering
	\includegraphics[width=10.5cm]{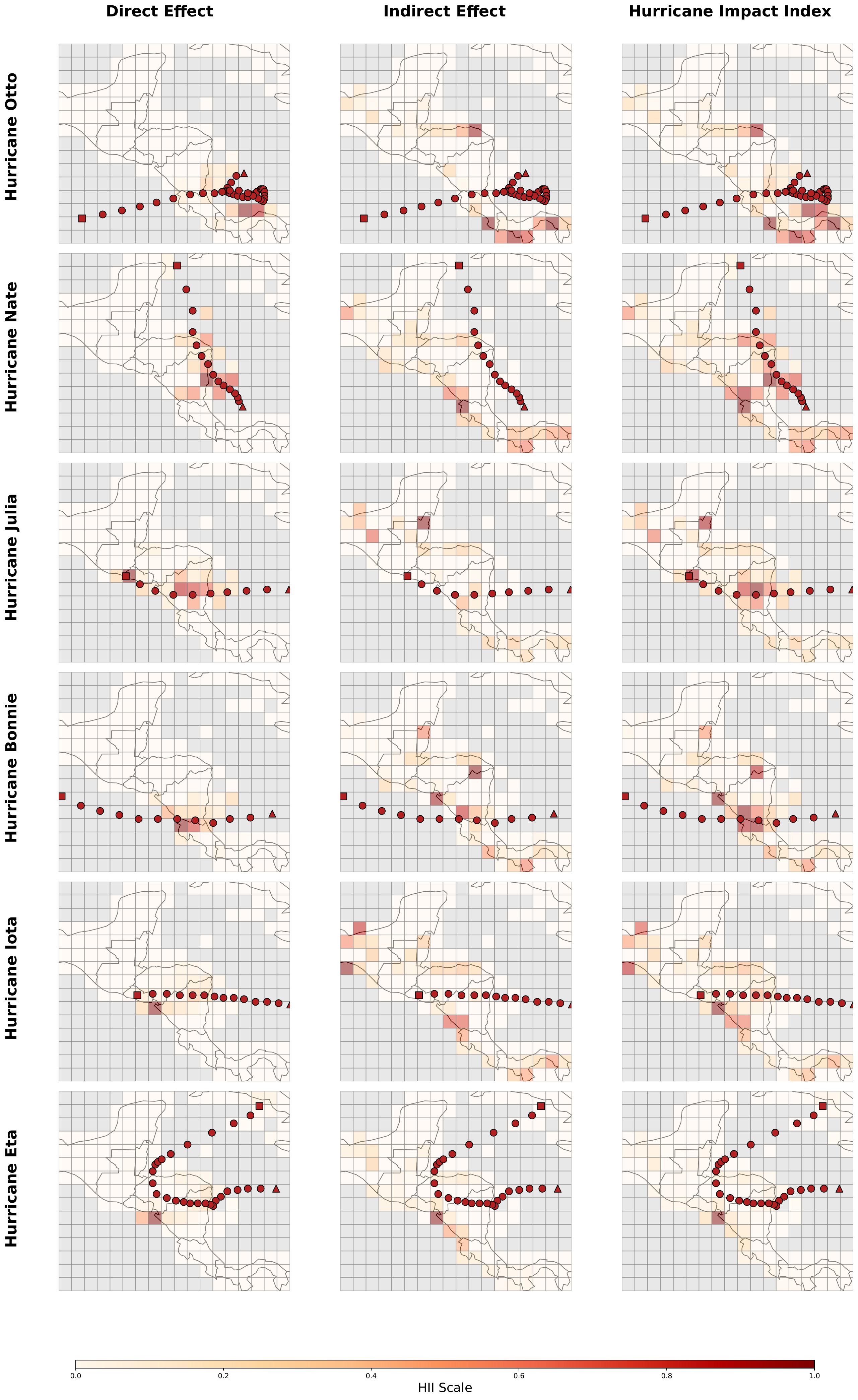}
	\caption{Direct Effect (First Column), Indirect Effect (Second Column), and HII (Third Column) for the Hurricane Tracks of Otto (2016), Nate (2017), Julia (2022), Bonnie (2022), Iota (2020), and Eta (2020), highlighting their respective impacts on the region, according to the formulas \eqref{eq:HII_direct}, \eqref{eq:HII_indirect} and \eqref{eq:HII_1}. The triangle indicates the hurricane's start, while the square marks its end.}
	\label{fig:HII_original}
\end{figure}

The results reveal notable variations in HII values across different hurricanes. Fig. \ref{fig:HII_original} shows both the segmented effects (see Equations \eqref{eq:HII_direct} and \eqref{eq:HII_indirect}, as well as the HII \eqref{eq:HII_1}), for each analyzed hurricane in different rows. The red dots represent the hurricane's position at specific points in space and time. Viewed together, they trace the hurricane's path. The triangle marks the starting point, while the square indicates its endpoint. We note that our method captures both the direct and indirect effects in a consistent manner with the region's orographic features, particularly for the indirect effect. In addition, the indicator is sufficiently flexible to account for the unique characteristics of each hurricane, which explains why the same grid cells do not necessarily activate across all analyzed events.

\begin{figure}
	\centering
	\includegraphics[width=13cm]{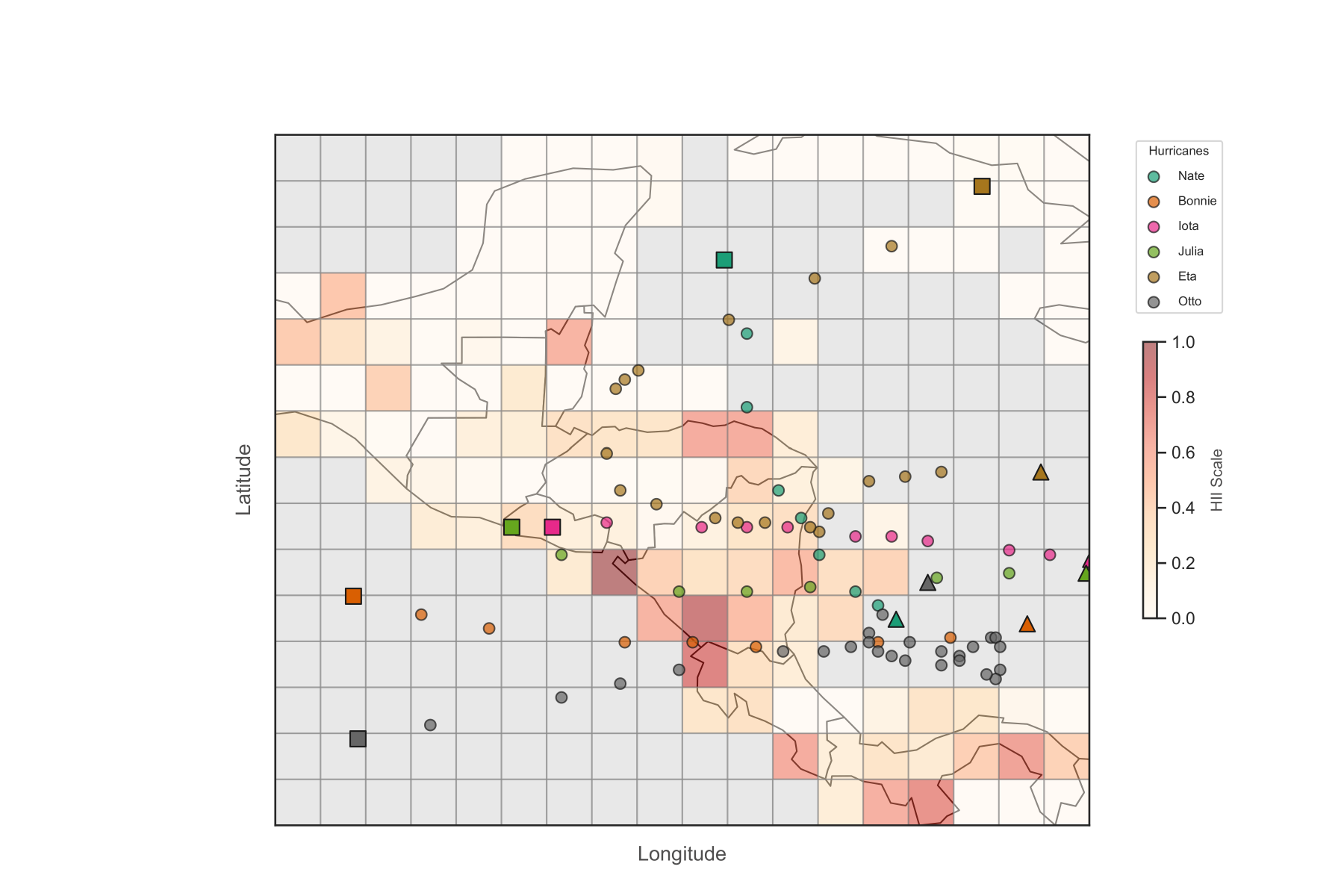}
	\caption{$\text{HII}_{R}$ for the Hurricanes Tracks of Otto (2016), Nate (2017), Julia (2022), Bonnie (2022), Iota (2020), and Eta (2020). The triangle indicates the hurricane's start, while the square marks its end.}
	\label{fig:HII_R}
\end{figure}

Fig. \ref{fig:HII_R} enables us to spatially aggregate the effects of the studied hurricanes using Equation \eqref{eq:HII_2}. In this case, HII$_R$ reflects how multiple events simultaneously affect specific grid cells, offering a broader perspective than evaluating events independently. Consequently, it highlights regions that experience more frequent impacts. As shown, the Pacific coast exhibits the highest index values, indicating that this coastline suffers greater impacts in Central America, while the aggregated impacts observed in the Caribbean predominantly correspond to direct effects.

\begin{figure}
	\centering
	\includegraphics[scale=0.25]{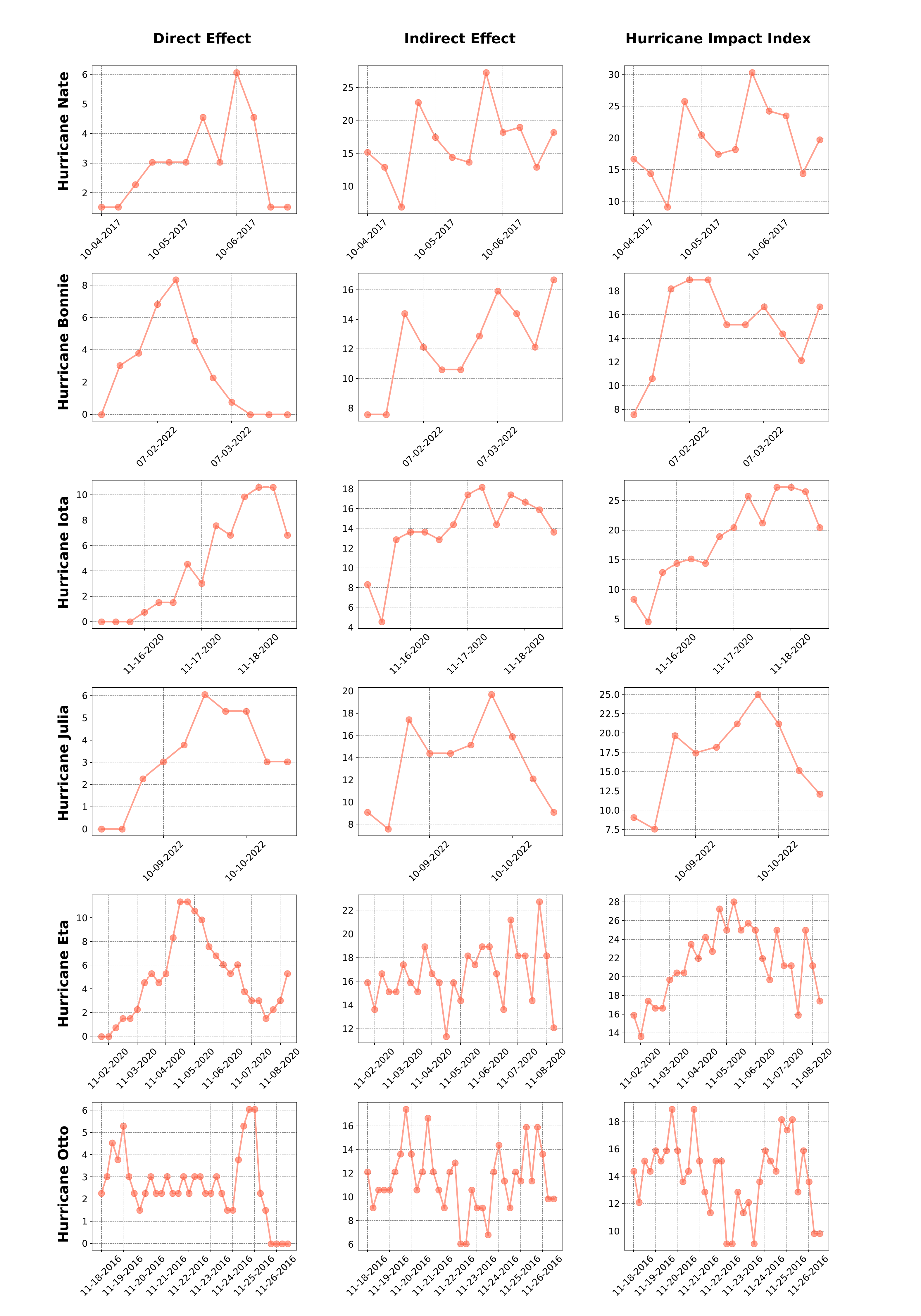}
	\caption{$\text{HII}_{\text{TS1},h,t}$  calculated using Equation \eqref{eq:HII_3}.}
	\label{fig:HII_TS2}
\end{figure}

Fig. \ref{fig:HII_TS2} illustrates the temporal evolution of the direct, indirect, and aggregated effects computed with Equation \eqref{eq:HII_3}. This indicator permits a dual comparison: (i) the absolute magnitude of each effect for a given event and (ii) the relative regional impacts across different events. For instance, Hurricane Nate produced an indirect impact roughly five times greater than its direct impact, whereas the direct–indirect balance was much closer for Hurricane Iota. Among the six hurricanes examined, Nate generated the most severe combined impact, while Otto and Bonnie produced the least.

Table \ref{tab:effect_peak} reports the timing of the peak normalized impact (Equation \eqref{eq:HII_4}) for the direct, indirect, and aggregated components. Across all events, the peaks occur within one day of one another, yet the corresponding Saffir–Simpson categories vary markedly. Hence, a tropical cyclone does not need to reach hurricane strength to inflict substantial regional damage.

\begin{table}
	\centering
	\begin{tabular}{c|c|c|c|c}
		\toprule
		Event & Direct  & Indirect & Combined & Category (Saffir-Simpson)\\ 
		\midrule
		Nate & October 6, 2017 & October 5, 2017 & October 5, 2017 & TD-TS\\
		Bonnie & July 02, 2022 & July 03, 2022  & July 02, 2022 & TS\\
		Iota & November 18, 2020 & November 17, 2020  & November 17, 2020 & C4\\
		Julia & October 09, 2022  & October 10, 2022  & October 10, 2022 & TS \\
		Eta & November 05, 2020 & November 06, 2020  & November 05, 2020 & TD\\
		Otto & November 24, 2016  & November 24, 2016  & November 24, 2016 & C2-C3\\
		\bottomrule
	\end{tabular}
	\caption{Dates when HII$_{TS2}$ peaked, separated by effect type, along with the corresponding observed Saffir–Simpson category at the point of maximum combined effect, using \cite{NOAA_HHT_2025}.}
	\label{tab:effect_peak}
\end{table}

\subsection{Comparative Analysis of Estimated and Observed Climate Event impacts}

To evaluate the reliability of the proposed index, we compared its estimated impacts with observed impacts derived from the DesInventar DataCards. Only the Costa Rica section was used, as it contains the sole continuous record for the years in which the hurricanes analyzed here occurred.

The DesInventar database is derived from news reports, which can introduce bias: media coverage typically concentrates on densely populated areas that experience pronounced damage. Consequently, sparsely populated regions—although affected—may be underreported, leading to their potential underrepresentation in DesInventar records.

\begin{figure}
	\centering
	\includegraphics[scale=0.25]{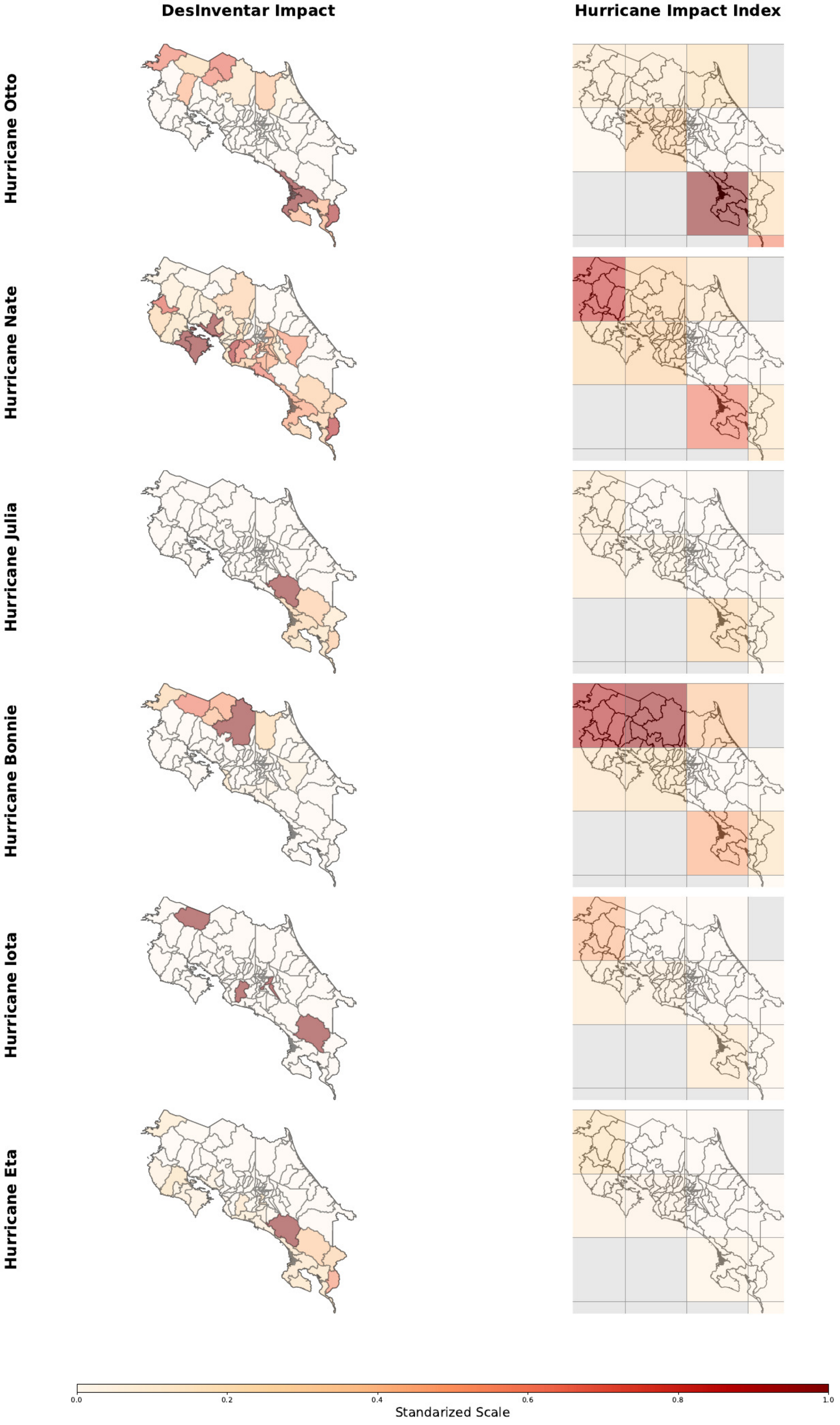}
	\caption{Comparison Between DesInventar DataCards Variable (left column) with the HII$_R$ (right column)}
	\label{fig:HIIRvsDesinventar}
\end{figure}

Across the panel (see Fig. \ref{fig:HIIRvsDesinventar}), we observe that HII$_R$ generally mirrors the spatial patterns of impact recorded in DesInventar. For Hurricane Otto, both sources highlight substancial impacts in northern Costa Rica and particularly in the southwestern region. Although this region does not fall within the highest category on the HII$_R$ color scale, the impact is still evident and consistent with the reported data. In the case of Hurricane Nate, the HII$_R$ highlights widespread coastal impact, especially along the western region. However, some areas in the center-north are less visible in the HII$_R$, likely due to the coarse $1^\circ \times 1^\circ$ grid resolution used in its calculation, which can mask smaller-scale events.

The panel also shows that the HII$_R$ aligns well with the DesInventar data for Hurricane Eta. While the HII$_R$ initially covers a broader area, filtering it to Costa Rica results in lower intensity values, which explains the lighter tones compared to the DesInventar map. Nevertheless, both sources reflect the event's substancial effect along the Pacific coast. Hurricane Iota's impact is represented more broadly in the HII$_R$ due to the large grid size, which results in more extensive high-impact areas. In contrast, DesInventar pinpoints specific locations, making the impact appear more localized. Despite this subtle scale differences, the general areas affected correspond well. Hurricane Bonnie reveals a mixed agreement between results. The HII$_R$ detects notable impacts in the northern region, which alings with DesInventar. However, it also shows elevated impact levels in some other areas not highlighted in DesInventar. This potential overestimation may result from unreported damages in remote or sparsely populated zones, which HII$_R$ still captures due to its use of environmental variables. 

Regarding Hurricane Julia, both sources indicate impacts in the southern region. Yet, the HII$_R$ appears to overestimate impact in adjacent coastal zones that are not reflected in DesInventar, likely due to the coarser spatial resolution of the index or the absence of officially reported damage in those areas. Overall, the panel illustrates that the HII$_R$ offers a valuable, climate-informed approximation of hurricane effects across Costa Rica, frequently aligning with the DesInventar records.

Additionally, correspondence tables with the HII were created, detailing the main impacts for the countries of Nicaragua, Panamá, El Salvador, Honduras y Guatemala, which can be found in the supplementary material in Tables S1, S2, S3, S4, and S5, respectively. These tables provide a comprehensive overview of the primary impacts observed in each country, facilitating the comparison between the estimated impacts generated by the proposed indicator and the reported impacts. This comparison analysis validates the reliability and applicability of HII across various geographical contexts within the study area, specifically the grid-level indicator HII$_R$.

\section{Conclusions and Discussion}

In this study, we designed a novel index that accounts for both the direct and indirect effects of hurricanes, applying it specifically to the Central America region. The HII effectively captures these effects in a consistent manner while remaining flexible enough to accommodate the unique characteristics of each hurricane. This adaptability is particularly important given the variability in hurricane paths, intensities, and local geographic factors that influence their impacts.

One of the key findings of our study was the appearance of higher index values along the Pacific coast, primarily due to indirect effects. The ability to aggregate HII values over time offers valuable insights for identifying areas with higher vulnerability to tropical cyclones. This spatial analysis can support decision-makers in prioritizing mitigation and preparedness strategies, ensuring that resources are allocated efficiently. Similarly, aggregating HII across space enables researchers to visualize the evolution of hurricane impacts throughout the region and identify periods of maximum intensity. Such insights are critical for recognizing conditions that contribute to the most severe damage and can inform early warning systems and response planning.

Another advantage of the HII is its applicability to simulations of future hurricane trajectories. By comparing projected hurricane impacts with historical data, this index provides a robust framework for assessing potential future threats over large geographic areas. The statistical reliability of the index ensures that many cyclonic events or realizations from model ensembles can be analyzed with relatively low computational cost. 
Despite its advantages, the validation of the HII was constrained by the lack of reliable quantitative data on past disaster events. However, qualitative information suggests that the index aligns well with observed patterns, even though this information may be subject to biases. In Costa Rica, where more comprehensive damage reports are available, the spatial distribution of official records closely matches the HII estimates, despite differences in spatial resolution. This agreement supports the credibility of the index as a valuable tool for impact assessment.
While we acknowledge the limitations of the HII, we believe it remains a useful and adaptable tool that can be applied to other regions facing hurricane hazards. Its relevance extends to disaster prevention and response organizations, municipalities, meteorological and hydrological services, and various governmental and non-governmental institutions. By providing a systematic approach to quantifying hurricane impacts, the HII contributes to more effective disaster preparedness and response strategies, ultimately enhancing resilience in vulnerable communities.

\subsection*{Acknowledgements}

HH wishes to acknowledge partial funding for this study through the following Vicerrectoría de Investigación, Universidad de Costa Rica grants: B9454 (supported by Fondo de Grupos), C2103, C3991 (UCREA), A4906 (PESCTMA), C5067, C3721, and B0-810. HH was partially supported by a grant awarded by the International Development Research Centre (IDRC), Ottawa, Canada, and the Central American University Council (CSUCA-SICA) to the Red Centroamericana de Ciencias sobre Cambio Climático (RC4) project (CR-66, C4468, SIA 0054-2, the opinions expressed here do not necessarily represent those of IDRC, CSUCA, or the Board of Governors). This work was partially produced when HH was on sabbatical leave from UCR. He thanks Vicerrectoría of Docencia for his research time allowance during 2024 and to the School of Physics during 2025. LAB gratefully acknowledges the support of the Vicerrectoría de Investigación at the Universidad de Costa Rica through grant C3197.

\section*{Declarations}


\begin{itemize}
	\item Funding
	
	The research leading to these results received funding from Vicerrectoría de Investigación, Universidad de Costa Rica under grant agreements B9454 (supported by Fondo de Grupos), C2103, C3991 (UCREA), A4906 (PESCTMA), C5067, C3721, and B0-810, and from International Development Research Centre (IDRC), Ottawa, Canada, and the Central American University Council (CSUCA-SICA) to the Red Centroamericana de Ciencias sobre Cambio Climático (RC4) project (CR-66, C4-468, SIA 0054-2). 
	
	\item Data availability
	
	All the sources of the data supporting the findings of this study are described and referenced in the Data subsection (\ref{Data}).
	
	\item Materials availability
	
	Not applicable.
	
	\item Code availability
	
	The code used for the calculations and generation of figures for this article is available at \href{https://github.com/ManriCamachoP/HII}{this repository}.
	
	\item Author contribution
	
	All authors contributed to the study conception, design, and analysis. The first draft of the manuscript was written by Manrique Camacho and Amanda Cedeño, and all authors commented on previous versions of the manuscript. All authors read and approved the final manuscript.
	
\end{itemize}

\bibliography{references}

\end{document}